\documentclass[twocolumn,showpacs,preprintnumbers,nofootinbib,amsmath,amssymb,floatfix,aps,prd,10pt]{revtex4-1}
\usepackage{graphicx}
\usepackage{subfigure}
\usepackage{isotope}
\usepackage{bm}

\DeclareMathOperator\artanh{artanh}

\graphicspath{{figs/}}

\begin{document}

\title{Improved estimate of the cross section for inverse beta decay }
\author{Artur M. Ankowski}
\email{ankowski@vt.edu}
\affiliation{Center for Neutrino Physics, Virginia Tech, Blacksburg, Virginia 24061, USA}

\date{\today}%

\begin{abstract}
The hypothesis of the conserved vector current, relating the vector weak and isovector electromagnetic currents,
plays a fundamental role in quantitative description of neutrino interactions. Despite being experimentally confirmed with great precision, it is not fully implemented in existing calculations of the cross section for inverse beta decay, the dominant mechanism of antineutrino scattering at energies below a few tens of MeV. In this article, I estimate the corresponding cross section and its uncertainty, ensuring conservation of the vector current.
While converging to previous calculations at energies of several MeV, the obtained result is appreciably lower and predicts more directional positron production near the reaction threshold. These findings suggest that in the current estimate of the flux of geologically produced antineutrinos the $^{232}$Th and $^{238}$U components may be underestimated by 6.1 and 3.7\%, respectively. The proposed search for light sterile neutrinos using a $^{144}$Ce--$^{144}$Pr source is predicted to collect the total event rate lower by 3\% than previously estimated and to observe a spectral distortion that could be misinterpreted as an oscillation signal. In reactor-antineutrino experiments, together with a re-evaluation of the positron spectra, the predicted event rate should be reduced by 0.9\%, diminishing the size of the reported anomaly.

\end{abstract}

%


\maketitle
Interactions of low-energy antineutrinos provide essential information on topics as seemingly distant as supernova explosions~\cite{Burrows:2000mk}, the energy budget of the Earth~\cite{Fiorentini:2007te}, neutrino oscillations~\cite{Vogel:2015wua}, and nuclear nonproliferation~\cite{Christensen:2013eza}.

Of utmost importance is the process of inverse beta decay (IBD),
\begin{equation}\label{eq:IBD}
\bar\nu_e+p\rightarrow e^++n,
\end{equation}
the cross section of which exceeds those for scattering off electrons and nuclei by a few orders of magnitude at energies below $\sim$20 MeV~\cite{Fogli:2004ff}.

At such kinematics, the estimate of the IBD cross section by Llewellyn Smith~\cite{LlewellynSmith:1971zm}---widely employed at higher energies---is not suitable because it neglects the difference between the neutron and proton masses. As a remedy, Vogel and Beacom~\cite{Vogel:1999zy} obtained a low-energy approximation of the cross section accounting for this difference, and analyzed the angular distribution of the produced positrons. As this approximation becomes inaccurate at energies above $\sim$20 MeV, Strumia and Vissani~\cite{Strumia:2003zx} performed fully relativistic calculations and compared them to existing theoretical results.



In this article, I point out that the vector part of the hadronic current employed in Refs.~\cite{Vogel:1999zy,Strumia:2003zx} is not conserved, although its conservation is invoked to express the vector form factors by their electromagnetic counterparts. I~remove this theoretical inconsistency by using an appropriate matrix representation of the current. This procedure sizably changes the description of the IBD process near the threshold, reducing the total cross section and increasing the directionality of the produced positrons.

The importance of these findings is demonstrated on the example of geoneutrinos~\cite{Araki:2005qa,Gando:2013nba,Agostini:2015cba} and the obtained results suggest that the current estimates of the flux may need to be revised, with particularly large increase of its thorium component. In addition, the reduced estimate of the cross section is able to explain part of the reactor-antineutrino anomaly~\cite{Mention:2011rk,Hayes:2013wra}, and its altered energy dependence is likely to affect the predicted spectra of produced positrons. An experimental verification of the predictions of this article---regarding both the reduced event rate and their energy distribution---may soon be provided by a sterile neutrino search using a $^{144}$Ce--$^{144}$Pr source~\cite{Cribier:2011fv,Borexino:2013xxa}.

To obtain the cross section, recall that the matrix element for IBD can be accurately calculated within the Fermi theory,
\begin{equation}
\mathcal M\propto J_\mu^\textrm{lept}\:J^\mu_\textrm{hadr},
\end{equation}
as an interaction between the leptonic and hadronic currents,
\begin{eqnarray}
J_\mu^\textrm{lept}&=&\bar u_e \gamma_\mu(1+\gamma_5) u_{\bar\nu},\\
J^\mu_\textrm{hadr}&=&\bar u_n (V^\mu+A^\mu) u_p,
\end{eqnarray}
with the Dirac spinors $u_{\bar\nu}=u_{\bar\nu}(k,\lambda)$, $u_e=u_e(k',\lambda')$, $u_p=u_p(p,s)$, and $u_n=u_n(p',s')$ describing the electron antineutrino, positron, proton, and neutron, respectively.

For the vector and axial parts of the hadronic current, $V^\mu$ and $A^\mu$, the matrix representations
\begin{eqnarray}
V^\mu_\textrm{std}&=&\gamma^\mu F_1+i\sigma^{\mu\kappa}\frac{q_\kappa}{2M}F_2,\label{eq:stdVectorCurrent}\\
A^\mu_\textrm{std}&=&\gamma^\mu\gamma_{5}F_A+\gamma_{5}\frac{q^\mu}M F_P,
\end{eqnarray}
with $M$ denoting the average nucleon mass, are widely adopted~\cite{LlewellynSmith:1971zm,Vogel:1999zy,Strumia:2003zx}. Note that second-class currents~\cite{Weinberg:1958ut} are disregarded in this work because the available experimental evidence points toward their nonexistence in nature~\cite{Sumikama:2008zz,Minamisono:2001cd,Minamisono:2011zz}.

Due to the isospin invariance of strong interactions, the vector currents $\bar u_n V^\mu u_p$ and $\bar u_p \gamma_0 V^{\mu\dagger} \gamma_0 u_n$ together with the isovector electromagnetic current form a triplet of conserved currents~\cite{Gershtein:1955fb,Feynman:1958ty}. As a consequence, the vector form factors $F_i$ ($i=1,2$) are related to the electromagnetic form factors of proton and neutron by
\begin{equation}
F_i=F_i^p-F_i^n.
\end{equation}
The numerical results presented in this article are obtained using
state-of-the-art parametrization of the nucleon form factors from Refs.~\cite{PhysRevC.70.068202,PhysRevLett.105.262302}.

Appearing in the axial part of the hadronic current, the pseudoscalar form factor $F_P$ can be expressed by the axial form factor $F_A$ as~\cite{PhysRevLett.4.380,Weinberg:1996kr}
\begin{equation}
F_P=\frac{2M^2}{m_\pi^2-q^2}F_A,
\end{equation}
where $m_\pi$ is the pion mass and $q=k-k'=p-p'$ denotes the four-momentum transfer.

The axial form factor, in turn, can be accurately parametrized as
\begin{equation}
F_A=\frac{g_A}{(1-q^2/M_A^2)^2},
\end{equation}
at the considered neutrino energies. The axial coupling constant $g_A=-1.2723(23)$~\cite{Agashe:2014kda} is extracted from neutron beta-decay measurements, and the axial mass $M_A=1.026(21)$ GeV is determined predominantly from neutrino scattering off deuteron~\cite{Bernard:2001rs}.

The hypothesis of the conserved vector current is experimentally confirmed with great precision~\cite{Severijns:2011zz,PhysRevC.91.025501}---at the level of $10^{-4}$---and plays fundamental role in the estimate of the IBD cross section. However, it is important to realize that the standard expression~\eqref{eq:stdVectorCurrent} violates current conservation when the difference between the neutron and proton masses is accounted for.
In fact,
\begin{equation}
q_\mu \bar u_n V^\mu_\textrm{std} u_p= \bar u_n q_\mu\gamma^\mu F_1 u_p=(M_n-M_p)F_1\bar u_n u_p,
\end{equation}
so the vector current is conserved only when the neutron and proton masses, $M_n$ and $M_p$, are considered to be equal.

At the kinematics of interest, conservation of the vector current (CVC) can be ensured by applying the modified matrix representation
\begin{equation}
V^\mu_\textrm{cvc}=\left(\gamma^\mu -\frac{q^\mu}{q^2}\gamma^\rho q_\rho\right)F_1+i\sigma^{\mu\kappa}\frac{q_\kappa}{2M}F_2.\label{eq:ConservedVectorCurrent}
\end{equation}
While this method has not been used in the previous estimates of the IBD cross sections~\cite{Vogel:1999zy,Strumia:2003zx}, it is known in the description of resonance excitation, see e.g.~\cite{Nowakowski:2004cv,PhysRevC.73.065502}.

The resulting differential cross section as a function of $q^2$ may be cast in the standard form~\cite{LlewellynSmith:1971zm,Strumia:2003zx}
\begin{equation}\label{eq:csc}
\frac{d\sigma^\textrm{tree}_\textrm{cvc}}{dq^2}=\frac{(G_F\cos\theta_C)^2}{8\pi M_p^2E_\nu^2}\big[M^4A+M^2B(s-u)+C(s-u)^2\big],
\end{equation}
where $s-u=4 M_pE_\nu+q^2-m^2-2M\Delta$, $E_\nu$ being the neutrino energy, and
\begin{widetext}
\begin{equation}\begin{split}\label{eq:ABC}
A&=4(\tau+\mu)\Big\{\left(\tau-\mu\right)\left[(F_1+F_2)^2+F_A^2\right]-(F_1-\tau F_2)^2 +F_A^2
+4\mu F_P(\tau F_P-F_A)\Big\}\\
&\quad-8\mu\frac{\Delta}{M}\left(F_1+F_2\right)F_A
+\frac{\Delta^2}{M^2}\left(\tau+\mu\right)\left[(F_1+F_2)^2-(1+\tau) F_2^2 -F_A^2+4\mu F_P^2\right]\\
&\quad+\frac{\Delta^2}{M^2}\left[\tau(F_1+F_2)^2-(1+\tau) F_1^2 -F_A^2 -4\mu F_PF_A - 2z \frac{\mu^2}{\tau}F_1F_2\right]\\
&\quad-z\frac{\Delta^2}{M^2}\left[\frac{\mu}{\tau}\left(1-\frac{\mu}{\tau}\right)+\mu\left(1+\frac{\mu}{\tau}\right)\right]F_1^2,\\
B&=4\tau\left(F_1+F_2\right)F_A-\mu\frac{\Delta}{M}\left[F_2^2+(1-z)F_1F_2+2F_AF_P+\frac{z}{\tau}F_1^2\right],\\
C&=\frac14\left(F_1^2+\tau F_2^2+F_A^2\right),
\end{split}\end{equation}
\end{widetext}
with $\tau=-q^2/4M^2$, $\Delta=M_n-M_p$, and $\mu=m^2/4M^2$. In numerical calculations, the state-of-the-art values are applied for the Fermi constant $G_F=1.1663787(6)\times10^{−5}$/GeV$^2$ and the cosine of the Cabibbo angle $\cos\theta_C=0.97425(22)$~\cite{Agashe:2014kda}.

The parameter $z$ in Eq.~\eqref{eq:ABC} is introduced to show the difference between the calculations with ($z=1$) and without ($z=0$) the restored conservation of the vector current. Note that in the latter case the obtained expressions reduce to those obtained by Strumia and Vissani~\cite{Strumia:2003zx}.

The kinematically allowed range of $q^2$ extends from $q^2_-$ to $q^2_+$ that can be expressed as
\begin{equation}
q^2_\pm=\frac{m^2M_p-E_\nu {\mathcal D}\pm E_\nu\sqrt{{\mathcal D}^2-4m^2M_n^2}}{M_p+2E_\nu}
\end{equation}
with
\begin{equation}
{\mathcal D}=(M_p+E_\nu)^2-E_\nu^2-M_n^2-m^2.
\end{equation}

To achieve the accuracy required by modern experimental applications, the radiative corrections to the tree-level cross section~\eqref{eq:csc} must be taken into account. Decomposing them into the inner and outer parts~\cite{Sirlin:1967zza,Wilkinson:1970}, as customary, and considering the leading order terms in the fine-structure constant $\alpha$, one obtains
\begin{equation}
\sigma_\textrm{cvc}=\int_{q^2_-}^{q^2_+} dq^2\frac{d\sigma^\textrm{tree}_\textrm{cvc}}{dq^2}\left[1+\delta_\textrm{in}+\delta_\textrm{out}(\beta)\right].
\end{equation}
For the inner correction, the value
\begin{equation}
\delta_\textrm{in}=0.02250(38),
\end{equation}
is adopted from the recent estimate~\cite{Marciano:2005ec}.
The outer correction $\delta_\textrm{out}(\beta)$ is a function of the positron's speed,
\begin{equation}
\beta=\sqrt{1-\frac{m^2}{E_e^2}}\quad\textrm{where}\quad E_e=E_\nu+\frac{q^2-2M\Delta}{2M_p},
\end{equation}
that can be expressed as~\cite{Fukugita:2004cq,Raha:2011aa}
\begin{equation}
\begin{split}
\frac{2\pi}{\alpha}\delta_\textrm{out}(\beta)&=\left(\frac{3\beta}{2}+ \frac{7}{2\beta}\right)\artanh\beta- \frac{8}{\beta}\artanh^2\beta\\
 &\quad+ \left( \frac{4}{\beta}\artanh\beta - 4 \right)\ln\frac{4\beta^2}{1 - \beta^2}\\
 &\quad+ \frac{8}{\beta}L\left( \frac{2\beta}{1 + \beta} \right)+ 3\ln\frac{M_p}{m} + \frac{23}{4},
\end{split}
\end{equation}
with
\begin{eqnarray}
\artanh x&=&\frac12\ln\frac{1+x}{1-x},\\
L(x)&=&\int_0^x dt\frac{\ln(1-t)}{t}.
\end{eqnarray}
Note that the above expressions are equivalent to those given in Ref.~\cite{Kurylov:2002vj}.

\begin{figure}
\centering
    \includegraphics[width=0.84\columnwidth]{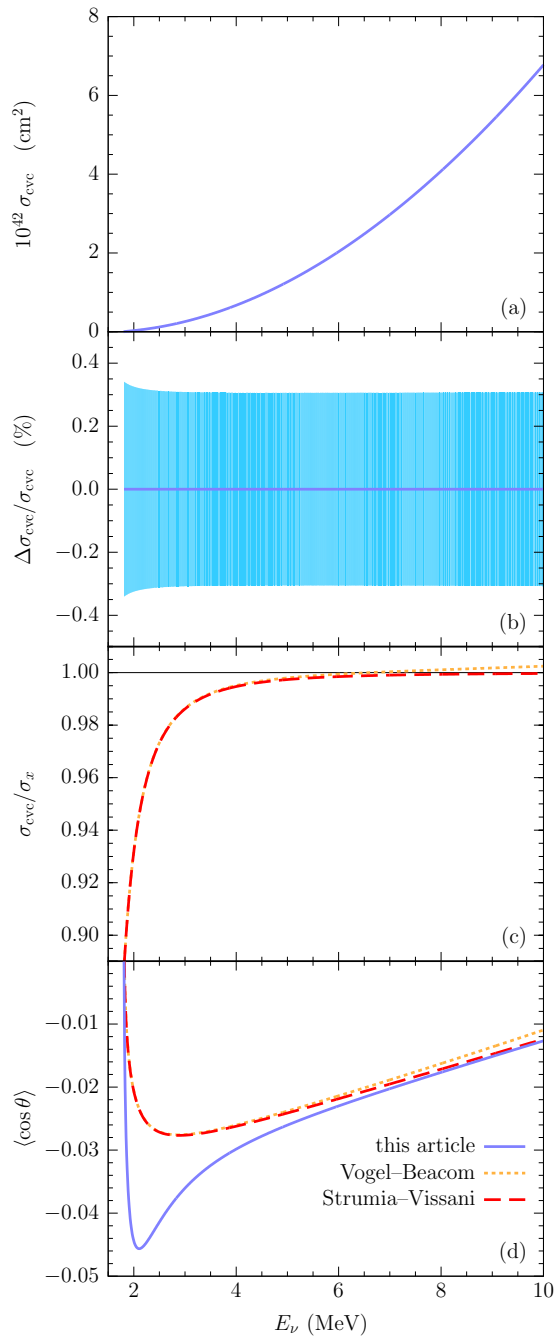}
    \subfigure{\label{fig:total}}
    \subfigure{\label{fig:uncertainty}}
    \subfigure{\label{fig:ratios}}
    \subfigure{\label{fig:cosTheta}}
\caption{\label{fig:totCS}(Color online). Estimate of (a) the IBD cross section $\sigma_\textrm{cvc}$ and (b) its uncertainty, obtained ensuring the conservation of the vector current. The relevance of this effect is shown by (c) the ratios of $\sigma_\textrm{cvc}$ to the results of Refs.~\cite{Vogel:1999zy,Strumia:2003zx} and (d) the comparison of the average cosine of the positron production angle calculated within these approaches.}
\end{figure}

The obtained total IBD cross section and its uncertainty are presented in Figs.~\ref{fig:total} and \ref{fig:uncertainty}, respectively. The uncertainty is estimated by varying parameters entering the calculations within their uncertainties and adding the corresponding cross section's variations in quadratures. The result---not exceeding 0.35\%---is overwhelmingly dominated by the uncertainty of the axial coupling constant, and contributions other then those from the inner radiative corrections and the Cabibbo angle can be safely neglected.

Playing an important role only at low absolute values of $q^2$, the CVC restoring procedure~\eqref{eq:ConservedVectorCurrent} affects the IBD cross section in an appreciable manner solely at low neutrino energies. This feature is illustrated in Fig.~\ref{fig:ratios}, showing the ratio of $\sigma_\textrm{cvc}$ to the calculations of Vogel and Beacom~\cite{Vogel:1999zy} and those of Strumia and Vissani~\cite{Strumia:2003zx}. In all the cases, the same treatment of the radiative corrections has been applied. While at $E_\nu=2$ MeV, the result of this article is lower than those of Refs.~\cite{Vogel:1999zy,Strumia:2003zx} by as much as $\sim$6.8\%, this effect reduces to $\sim$0.5\% at 4 MeV. Note that at higher energies the difference between the cross sections of Refs.~\cite{Vogel:1999zy} and~\cite{Strumia:2003zx} gradually becomes visible but remains below 0.5\% for $E_\nu\leq13$ MeV.

As the kinematic region of low $|q^2|$ corresponds to high $\cos\theta$, with $\theta$ being the positron's production angle, the observed reduction of the cross section at low $|q^2|$ translates into a decrease of the average value of $\cos\theta$, shown in Fig.~\ref{fig:cosTheta}. The manifest increase of the directionality at energies $E_\nu\sim2$--3 MeV, resulting predominantly from the last term in the $B$ factor~\eqref{eq:ABC}, may be relevant, e.g., for spatially mapping geoneutrinos~\cite{Safdi:2014hwa}.

To obtain the $\langle\cos\theta\rangle$ dependence on antineutrino energy presented in Fig.~\ref{fig:cosTheta}, the $\cos\theta$-even and $\cos\theta$-odd parts of the cross section has been treated separately, as they are subject to different outer radiative corrections~\cite{Fukugita:2004cq,Raha:2011aa}. However, in agreement with the conjecture of Ref.~\cite{Vogel:1999zy}, this procedure turns out to have a small effect on the result, affecting it by no more than 0.15\% for $E_\nu\geq2$ MeV.

The IBD cross section is generally considered to be subject to low uncertainties and, therefore, its CVC-related reduction may have important consequences. For example, in the context of a determination of the geoneutrino flux, I find that $\sigma_\textrm{cvc}$ leads to the $^{232}$Th and $^{238}$U components higher by 6.1 and 3.7\%, respectively, than the estimates based on the cross sections of Refs.~\cite{Vogel:1999zy,Strumia:2003zx}. Those values are calculated using the spectra of Ref.~\cite{Enomoto:2006} and refer to the KamLAND site.

In Ref.~\cite{Cribier:2011fv}, it has been proposed to search for light sterile neutrinos using a $^{144}$Ce--$^{144}$Pr $\bar\nu_e$ source. Using the flux of Ref.~\cite{Gaffiot:2014aka}, I predict an overall 3\% reduction of the event rate with respect to simulations employing the cross sections of Refs.~\cite{Vogel:1999zy,Strumia:2003zx}, and a spectral distortion that may mimic an oscillation signal. As such an experiment is currently underway~\cite{Borexino:2013xxa}, these predictions can be tested within a 1-year time frame.

The reevaluation of the antineutrino spectra emitted by nuclear reactors~\cite{Mueller:2011nm} has recently lead to the conclusion that the event rates observed in past reactor experiments underestimate the predicted rates by $5.7\pm2.3$\%~\cite{Mention:2011rk}. Combining the contributions from the individual isotopes~\cite{Mueller:2011nm,Huber:2011wv} according to the weights~\cite{An:2015nua}, I estimate that the CVC-related reduction of the cross section lowers the predicted rate by 0.9\%, reducing the reactor anomaly.

Moreover, as the antineutrino energy is closely related to the prompt energy of the produced positron,  $E_\textrm{prompt}\simeq E_\nu - 0.78$ MeV, the results of Fig.~\ref{fig:ratios} corresponding to the low-$E_\textrm{prompt}$ region can be expected to bring into better agreement the predictions and the prompt energy spectra measured in near detectors of ongoing reactor experiments~\cite{An:2015nua,RENO:2015ksa}.

In summary, conservation of the vector part of the weak current has important consequences for the description of inverse beta decay at the kinematics corresponding to low $|q^2|$. For energies in the vicinity of the threshold, this effect sizably lowers the total cross section and increases the directionality of positron production. These results, of particular relevance for an estimate of the geoneutrino flux, may soon be verified by an experiment employing a $^{144}$Ce--$^{144}$Pr source to search for light sterile neutrinos. As a consequence of the reduced total cross section, the size of the reactor anomaly is also diminished, and the agreement between the predicted and observed positron spectra can be expected to improve in reactor-antineutrino experiments.

To facilitate analysis of other implications of the IDB cross section reported here---for example in the context of big-bang nucleosynthesis---its \textsc{c++} implementation, approximated expressions valid at low energies, off-shell generalization, and tabulated values are provided in Supplemental Material~\cite{SupplementalMaterial}.

\begin{acknowledgments}
I am indebt to Makoto Sakuda for drawing my attention to neutrino interactions at low energies. Special thanks are addressed to Patrick Huber and Jonathan Link for mostly informative discussions on the reactor-antineutrino anomaly and their valuable suggestions on the composition of this article. My work was supported by the National Science Foundation under Grant No. PHY-1352106.
\end{acknowledgments}


%

\end{document}